# Microsecond electro-optic switching in the nematic phase of a ferroelectric nematic liquid crystal

KAMAL THAPA,[1,2] SATHYANARAYANA PALADUGU,[1] AND OLEG D. LAVRENTOVICH[1,2,3*]

[1]*Advanced Materials and Liquid Crystal Institute, Kent State University, Kent, OH 44242, USA*
[2]*Department of Physics, Kent State University, Kent, OH, 44241, USA*
[3]*Materials Science Graduate Program, Kent State University, Kent, OH 44242, USA*
*\*olavrent@kent.edu*

**Abstract:** Nematic liquid crystals exhibit nanosecond electro-optic response to an applied electric field which modifies the degree of orientational order without realigning the molecular orientation. However, this nanosecond electrically-modified order parameter (NEMOP) effect requires high driving fields, on the order of $10^8$ V/m for a modest birefringence change of 0.01. In this work, we demonstrate that a nematic phase of the recently discovered ferroelectric nematic materials exhibits a robust and fast electro-optic response. Namely, a relatively weak field of $2 \times 10^7$ V/m changes the birefringence by $\approx 0.04$ with field-on and-off times around 1 μs. This microsecond electrically modified order parameter (MEMOP) effect shows a greatly improved figure of merit when compared to other electro-optical switching modes in liquid crystals, including the conventional Frederiks effect, and has a potential for applications in fast electro-optical devices such as phase modulators, optical shutters, displays, and beam steerers.

## 1. Introduction

Nematic liquid crystals (NLCs) are widely used in electro-optic applications because of their intrinsic anisotropy of dielectric and optical properties [1]. Most applications rely on the Frederiks transition [2], in which an external electric field reorients the direction of average molecular orientation, called the director $\hat{\mathbf{n}}$. The detrimental feature of the Frederiks effect is its slow field-off relaxation, on the order of milliseconds [3]. When the field is switched off, the elastic nature of an NLC forces $\hat{\mathbf{n}}$ to return to its original orientation set up by the surface anchoring forces. The relaxation time $\tau_{\text{off}}$ of this passive process is determined by the rotational viscosity $\gamma_1$, elastic costant $K$, and thickness $d$ of the cell, $\tau_{\text{off}} = \frac{\gamma_1 d^2}{\pi^2 K}$. For a typical NLC, $\gamma_1 = 0.1$ kg/m $\cdot$ s, $K = 10$ pN and $d = 5$ μm, one finds $\tau_{\text{off}} \approx 25$ ms [3]. Many approaches have been developed over the years to speed up the relaxation; among these are the optimization of viscosity and elastic constants, development of the so-called dual frequency NLCs [3-8], polymer-layer-free alignment [9], design of gradient electric fields [10], and doping NLCs with various additives [11]. Still, the relaxation time of $\hat{\mathbf{n}}$ during the field-off state is in the millisecond or sub-millisecond range.

   A different approach to the electro-optic switching is offered by the so-called nanosecond electrically-modified order parameter (NEMOP) effect, in which the applied electric field modifies the degree of orientational order but does not alter the orientation of $\hat{\mathbf{n}}$ [12-17]. The NEMOP effect was first demonstrated for an NLC CCN-47 with a negative dielectric anisotropy $\Delta \varepsilon = \varepsilon_\parallel - \varepsilon_\perp \approx -5$ [12, 14]; here the subscripts refer to the direction parallel and perpendicular to $\hat{\mathbf{n}}$, respectively. An electric field on the order of $10^8$ V/m applied perpendicularly to $\hat{\mathbf{n}}$ in a planar cell, causes a uniaxial-to-biaxial transformation of the order parameter and changes birefringence by $\approx 0.001$ with the field-on $\tau_{\text{on}}$ and -off $\tau_{\text{off}}$ times being on the order of 10 ns [12, 14]. The reason for such a fast nanosecond switching is that the orientational order is altered at the molecular scale and does not involve collective realignment. Since the NEMOP effect is essentially a molecular-scale phenomenon, the amplitude and the relaxation times of birefringence change depend strongly on the details of molecular structure. In particular, an order-of-magnitude enhancement of the birefringence change up to ≈ 0.01 was achieved in nematics HNG 715600-100 and HNG 705800-100 with negative dielectric anisotropy, $\Delta\varepsilon \approx -10$, stronger than that of CCN-47 [13], and up to ≈ 0.02 when doped with highly polar molecules carrying a large transversal dipole [16]. The approach to modify the degree of orientational order can also be applied to nematics with $\Delta\varepsilon > 0$ in a homeotropic cell, by applying the field along $\hat{\mathbf{n}}$ and using oblique incidence of light [15, 18]. In particular, a nematic GPDA200 with $\Delta\varepsilon \sim 100$ demonstrated a strong birefringence change ≈ 0.04 switched by a moderate field of $3 \times 10^7$ V/m, with $\tau_{on} \approx \tau_{off} \approx 10$ μs [18]. In the case of $\Delta\varepsilon > 0$, the birefringence change is caused not only by the modification of the order parameter but also by the quenching of director fluctuations. The quenching is slower than the field-induced uniaxial-biaxial transformation. These prior experiments suggest that the performance of the electro-optical response of NLCs can be improved through searching for new materials with a molecular structure that favors a fast and large field-induced change of birefringence. Guided by this idea, in this work we explore the electro-optical switching of the nematic phase of ferroelectric nematic materials.

Ferroelectric nematic materials have been synthesized and characterized only recently [19-24]. These materials are formed by rod-like molecules with large longitudinal electric dipoles ~10 D. The molecules align either in a paraelectric fashion, forming a conventional paraelectric nematic (N), or in a polar fashion, with all the dipoles pointing in the same direction, forming a ferroelectric nematic ($N_F$) with a macroscopic spontaneous polarization **P** along $\hat{\mathbf{n}}$. The strong polarity of the molecules yields a large $\Delta\varepsilon > 0$ [25], and strong birefringence [23, 26, 27] in both N and $N_F$ phases. We demonstrate that the paraelectric nematic of the ferroelectric nematic mixture FNLC919 [26] with a large positive dielectric anisotropy $\Delta\varepsilon \approx (90 - 190)$ shows a strong change of birefringence, ≈ 0.04 when a relatively weak electric field ≈ $2 \times 10^7$ V/m is applied parallel to $\hat{\mathbf{n}}$. The response is fast, with $\tau_{on} \approx \tau_{off} \approx 1$ μs. The performance of different electro-optical modes and materials can be compared by a commonly used figure of merit (FoM) = $\delta\Gamma_{max}^2/\pi^2\tau_{off}$ [3, 28], where $\delta\Gamma_{max} = d \times \delta n_{max}$ and $\delta n_{max}$ is the amplitude of the field-induced birefringence . For the conventional Frederiks transition, the maximum possible $\delta n_{max}$ is the material birefringence $\Delta n = n_e - n_o \approx 0.2$, $\tau_{off} \approx 25$ ms, so that for $d = 5$ μm, one finds FoM ≈ 4 μm$^2$/s; here $n_e$ and $n_o$ are the extraordinary and ordinary refractive indices, respectively. The material explored in this work yields a much larger FoM ≈ $10^3$ μm$^2$/s. The latter is also more than an order of magnitude higher than the FoM reported for a paraelectric N material GPDA200 with $\Delta\varepsilon \sim 100$ [18], thanks to a 10 times faster switching speed.

## 2. Materials and methods

### 2.1 Dielectric and optical properties of the nematic phase

The explored material FNLC919 (Merck KGaA) [26], shows the following phase sequence upon cooling from the isotropic (I) phase: I - 82 °C - N - 46 °C - $N_X$ - 32 °C - $N_F$, where $N_X$ is an intermediate phase between the paraelectric N and the ferroelectric $N_F$ phases. The frequency-dependent dielectric permittivities of the N phase are determined by measuring the capacitance of cells with a thickness of $d = 55$ μm using an impedance analyzer (SI-1260, Schlumberger) over a temperature range of $(70 - 80)$ °C and a frequency range of 100 Hz - 10 MHz. The dielectric permittivity $\varepsilon_\parallel$ along $\hat{\mathbf{n}}$ is measured in a cell with the homeotropic alignment of $\hat{\mathbf{n}}$, set by spin-coated layers of polyimide SE5661 (Nissan Chemicals, Ltd.). The measured $\varepsilon_\parallel$ is large and highly dependent on both frequency and temperature, exhibiting a single Debye relaxation time $\tau_D$, which is in the range (6-15) µs, depending on the temperature in the N phase, Fig. 1(a). The perpendicular component, $\varepsilon_\perp$, Fig. 1(b), is measured in a planar cell with a rubbed PI2555 coating. The significantly larger parallel component $\varepsilon_\parallel$ compared to the perpendicular component $\varepsilon_\perp$, produces a large positive dielectric anisotropy $\Delta\varepsilon = \varepsilon_\parallel - \varepsilon_\perp$, which increases upon cooling from about 90 at 80°C to about 190 at 70°C, Fig. 1(c). Temperature dependence of birefringence $\Delta n$, Fig. 1(d), is measured at the wavelength $\lambda = 632.8$ nm by an interferometric technique using a wedge cell [29] with a small wedge angle of 0.806°. The wedge cell is assembled from two glass substrates coated with the polyimide PI2555 (HD MicroSystems) and unidirectionally buffed perpendicularly to the thickness gradient [30] to provide a homogeneous planar alignment and avoid a geometrical anchoring effect [31]. The thin part of the wedge is glued by a UV curable glue NOA68 (Norland Products, Inc.) without spacers, while the thickest part contains soda lime glass microspheres (Cospheric LLC) of a diameter 210 µm mixed with NOA68.

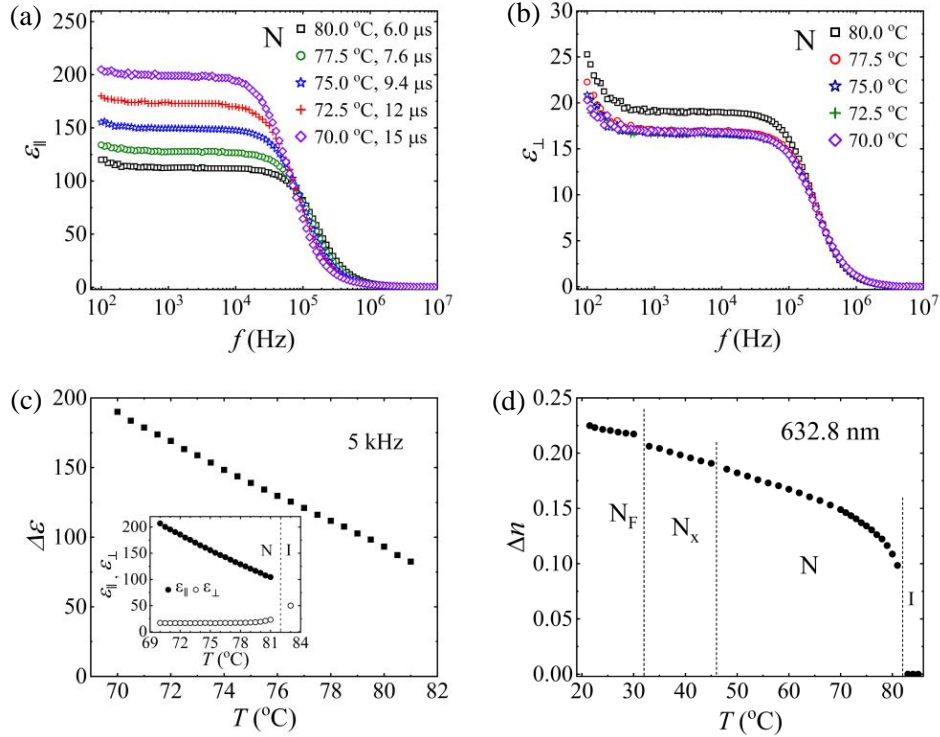

Fig. 1 Material parameters of FNLC919. Dielectric properties at various temperatures in the high-temperature N phase: (a) Frequency dependence of $\varepsilon_\parallel$ and the corresponding Debye relaxation times; (b) Frequency dependence of $\varepsilon_\perp$; (c) Temperature dependence of $\Delta\varepsilon$ and dielectric permittivities at 5 kHz. (d) Temperature dependence of birefringence at $\lambda = 632.8$ nm in different phases.

*2.2 Experimental setup and method of electro-optic switching of the nematic phase*

The electro-optical response is studied in the temperature range $T = (70 - 80)$ °C by applying a short voltage pulse to a flat N slab of a thickness $d = 5.2$ µm. The cell is assembled from two indium-tin-oxide (ITO) patterned glass substrates, spin-coated with SE5661. The transparent patterned ITO electrodes are of a low sheet resistance of 10 Ω/□ and a small active area of 2 mm × 2 mm. The layer of SE5661 sets homeotropic alignment in the N phase, Fig. 2(a), allowing us to apply the electric field along $\hat{\mathbf{n}}$. Since $\Delta\varepsilon > 0$, the electric field modifies the degree of orientation order without reorientation of the director $\hat{\mathbf{n}}$. The field-induced birefringence $\delta n$ is measured for an oblique 45° incidence of a He-Ne laser beam, $\lambda = 632.8$ nm, which passes through the LC cell and a Soleil-Babinet compensator, both in between two linear polarizers, Fig. 2(b). The N cell is placed between two prisms inside a custom-made copper holder. The projection of $\hat{\mathbf{n}}$ onto the plane of the polarizer and analyzer and the slow axis of the compensator are parallel to each other and make an angle of 45° with both polarizing directions. The N cell, prisms, and copper holder are thermally insulated inside a custom-made Teflon holder. The transmitted light intensity is measured using a photodetector TIA-525 (Terahertz Technologies, response time < 1 ns). Input voltage pulses of a duration 8 μs are applied by a system of waveform generator WFG500 (FLC electronics), high direct current (DC) voltage source KEITHLEY 237 (Keithley), and pulse generator HV 1000 pulser (Direct Energy). The input voltage pulse and output photodetector signal are monitored and recorded by a digital oscilloscope TDS2014 (Tektronix), sample rate 1GS/s. The temperature of the cell is controlled by a hot stage LTS120 and a controller PE94 (both Linkam) with an accuracy of 0.01°C.

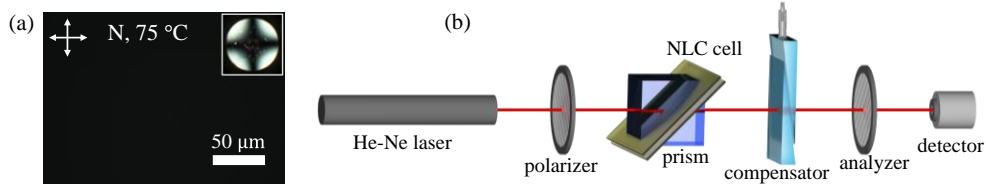

Fig. 2 (a) Polarizing optical microscopy texture of a homeotropic cell ($d = 5.2$ μm) in the N phase (75 °C) with crossed polarizers; inset shows the conoscopy image. (b) Experimental setup.

The dynamics of the field-induced birefringence $\delta n(t)$ is calculated using the four-point measurement scheme, described in Ref. [18]. The technique uses four successive measurements of the transmitted intensities $I_k$, $k = 1, 2, 3, 4$, subjected to an identical electrical pulse at four different settings of the retardance $\Gamma_k^{SB}$ of Soleil-Babinet compensator:

$$I_k(t) = [I_{\max}(t) - I_{\min}(t)] \sin^2\left\{\frac{\pi}{\lambda}\Gamma_k(t)\right\} + I_{\min}(t), \quad (1)$$

where $\Gamma_k(t) = \Gamma_N(0) + \Gamma_k^{SB} + \delta\Gamma(t)$ is the total dynamic optical retardance of the system and $\Gamma_N(0)$ is the optical retardance of the N cell when there is no electric field. We select $\Gamma_k^{SB} = \lambda(m + k/4) - \Gamma_N(0)$, where $m$ is an integer, in such a way that the initial intensities with zero field are maximum $I_2(0)$, minimum $I_4(0)$, and mean values $I_1(0) = I_3(0) = [I_2(0) + I_4(0)]/2$. For this selection, the field-induced birefringence is calculated as

$$\delta n(t) = \frac{\lambda}{2\pi d}\arg\{I_2(t) - I_4(t) + i[I_1(t) - I_3(t)]\}. \quad (2)$$

The switching-on $\tau_{\text{on}}$ and switching-off $\tau_{\text{off}}$ times are defined as the times within which $\delta n(t)$ changes between 10% and 90% of its maximum value $\delta n_{\max}$.

### 3. Results and discussion

The microsecond dynamics of the field-induced birefringence $\delta n(t)$ of the N cell is shown in Fig. 3(a). The birefringence amplitude $\delta n_{\max}$ increases with the driving field, Fig.3 (b,d), and the temperature, see Fig. 3(c) and the inset in Fig. 3(d). The switching times $\tau_{\text{on}}$ and $\tau_{\text{off}}$ are around 1 μs; the lowest value of the sum $\tau_{\text{on}} + \tau_{\text{off}} \approx 1.7$ μs is achieved at 78 °C and $E = 19$ V/μm, Fig. 3(e,f).

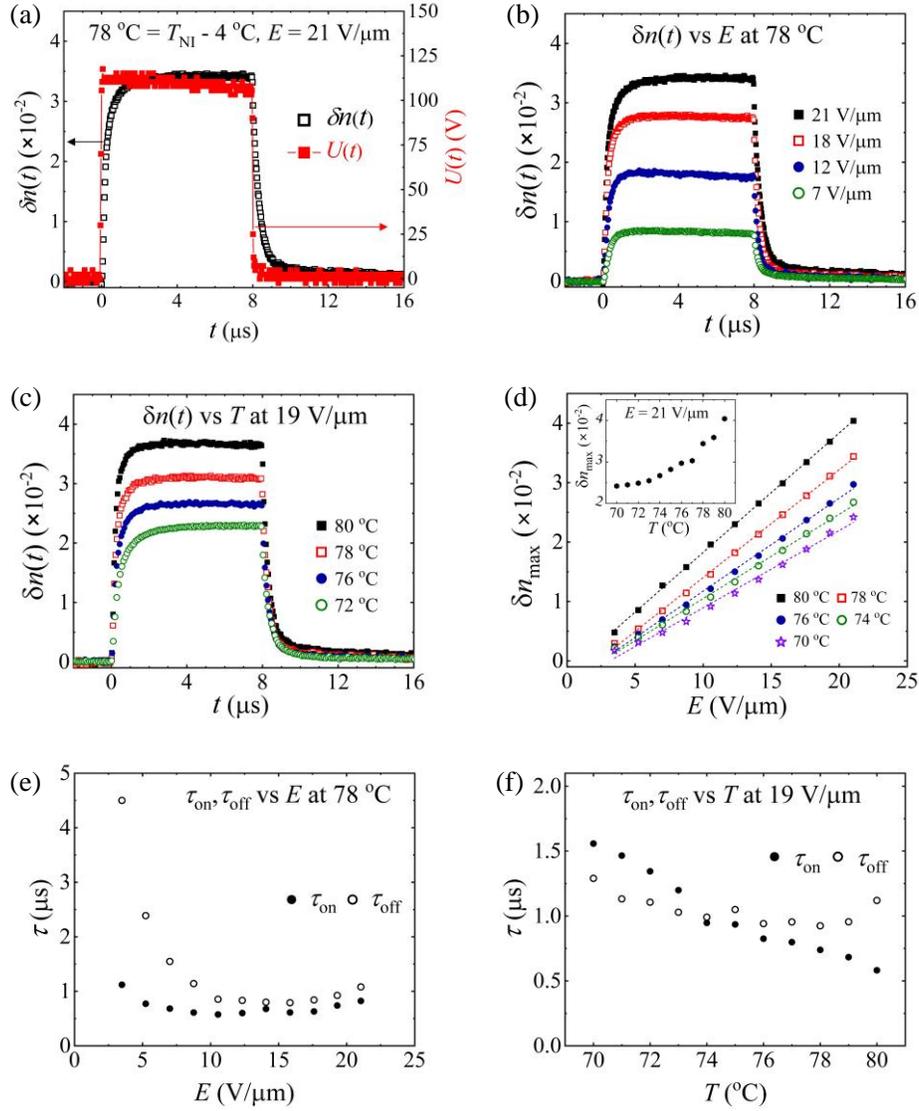

Fig. 3 Microsecond electro-optic switching of birefringence in the N phase of the ferroelectric FNLC919. (a) Dynamics of birefringence $\delta n(t)$ in response to the applied voltage pulse $U$ at 78 °C; (b) dynamics of $\delta n(t)$ at various voltages, 78 °C; and (c) at various temperatures, $E =$ 19 V/μm; (d) Amplitude $\delta n_{max}$ of field-induced birefringence as a function of electric fields at various temperatures with the dashed lines showing the linear fittings and the inset image showing the temperature dependence of $\delta n_{max}$ at 21 V/μm. Switching times at various (e) electric fields at 78 °C and (f) temperatures at 19 V/μm.

Table 1 compares the electro-optical performance of the studied $N_F$-forming material to the previously reported materials that do not form the $N_F$. The table shows the amplitude $\delta n_{max}$ of the field-induced birefringence, switching speeds, required driving electric fields, and FoMs. In terms of the birefringence change, a close performance was recorded for a paraelectric N material GPDA200 [18] which yields $\delta n_{max} \approx 0.04$ achieved at $E \approx 3 \times 10^7$ V/m with switching times $\approx 10$ μs. In our FNLC919 material, a similar $\delta n_{max} \approx 0.04$ is obtained at a somewhat lower field $E \approx 2 \times 10^7$ V/m, but with much shorter switching times $\approx 1$ μs. In the enhanced NEMOP [16], switching times are even shorter $\approx 0.1$ μs, but the birefringence change is weaker, $\delta n_{max} \approx 0.02$, while the driving field is higher, $E \approx 2 \times 10^8$ V/m.

Table 1. Field-induced Birefringence in Different Electro-optic Effects that Do Not Involve Director Realignment; NEMOP Stands for Nanosecond Electric Modification of the Order Parameter; MEMOP Stands for Microsecond Electric Modification of the Order Parameter.

| Effect | Material | $\Delta\varepsilon$ | $T$ (°C) | $E$ (V/m) | $\delta n_{max}$ | $\tau_{on}$ | $\tau_{off}$ | FoM[a] (μm²/s) |
|---|---|---|---|---|---|---|---|---|
| NEMOP [13] | HNG715600-100 | -12 | 23 | $1.7 \times 10^8$ | 0.013 | 30 ns | 30 ns | $1.3 \times 10^4$ |
| Enhanced NEMOP [16] | HNG715600-100 doped with DPP (19 wt.%) | -14 | 23 | $1.9 \times 10^8$ | 0.020 | 130 ns | 70 ns | $1.4 \times 10^4$ |
| MEMOP[18] in the paraelectric N material | GPDA200 | 30 at 70 °C | 72 | $2.8 \times 10^7$ | 0.038 | 7 μs | 13 μs | $2.8 \times 10^2$ |
| MEMOP in the N phase of the $N_F$ forming material | FNLC919 | 90 at 5 kHz | 80 | $2.1 \times 10^7$ | 0.040 | 0.65 μs | 1.3 μs | $3.0 \times 10^3$ |

[a]FoMs are calculated by taking cell thickness $d = 5$ μm

In electrically controlled liquid crystal-based optical retarders, the goal is to switch a sufficiently large optical retardance $\delta\Gamma_{max}$, at least equal to half of the probing wavelength $\lambda/2$ within the shortest possible time. These features are characterized by FoM expressed as $\delta\Gamma_{max}^2/\pi^2\tau_{off}$. The FOM of FNLC919 is on the order of $10^3$ μm²/s ($\delta n_{max} = 0.04, d \approx 5$ μm, $\tau_{off} \approx 1$ μs), which is one order of magnitude higher than the FoM of GPDA200 (FOM$\sim 10^2$ μm²/s, $\delta n_{max} = 0.04, d \approx 5$ μm, $\tau_{off} \approx 10$ μs) [18], and is at least two to three orders of magnitude higher than the FoM of the conventional Frederiks effect, $\sim (1-10)$ μm²/s [3, 28, 32]. Importantly, the demonstrated $\delta n \approx 0.04$ of the explored FNLC919 allows one to achieve a reasonably large retardance $\delta\Gamma \sim 400$ nm in relatively thin 10 μm cells.

In NLCs, the field-induced birefringence $\delta n$ is caused by the modification of the uniaxial order parameter [14] and quenching of the director fluctuations [32, 34]. The modification of the uniaxial order parameter is proportional to the square of an electric field [14]. In our case, $\delta n_{max}$ grows linearly with the field in the range $(3-21)$ V/μm, Fig. 3(d), indicating that the quenching of director fluctuations dominates over the modification of the uniaxial order parameter, similar to the previously reported effects in GPDA200 [18] and in 5CB [34]. As discussed in Ref. [18], the predominance of fluctuation quenching in a homeotropic N cell with $d = 5$ μm and $\Delta\varepsilon = 100$, sets in at the fields above $E_{fluct} = \beta/d\sqrt{\Delta\varepsilon} \approx 2$ V/μm, where $\beta \sim 100$ V is parameter deduced from the numerical fitting of 5CB behavior. The field-induced birefringence change caused by the quenching of director fluctuations has been calculated as [14, 18],

$$\delta n_{max} \approx \frac{6k_B T \Delta n \sqrt{\varepsilon_0 \Delta\varepsilon}}{\sqrt{2} K_{eff}^{3/2}} E, \qquad (3)$$

where $T$ is the absolute temperature, $k_B = 1.38 \times 10^{-23}$ J/K is the Boltzmann constant, $\varepsilon_0 = 8.85 \times 10^{-12}$ F/m is the dielectric permittivity of free space, and $K_{eff} = (4K_3)^{1/3}/(K_1^{-1} + K_2^{-1})^{2/3}$ is the effective elastic constant with $K_1, K_2$, and $K_3$ being the splay, twist, and bend elastic constants of the N, respectively. The last formula suggests that the smaller driving field causes the same $\delta n_{max}$ in FNLC919 as compared to GPDA200 in Ref. [18], which can be attributed to a larger $\Delta\varepsilon$ of FNLC919. The data presented in Ref. [18] for GPDA200 with $\Delta\varepsilon \approx 30$ at about 2 °C below the N-I phase transition yield the linear slope $\delta n_{max}/E \approx 0.14$ μm/V. In FNLC919 with $\Delta\varepsilon \approx 90$ at the similar distance from the NI transition point, this slope is steeper, $\delta n_{max}/E \approx 0.20$ μm/V, Fig. 3(d). The ratio of slopes is $0.14/0.20 \approx 0.7$ which is close to the dielectric ratio $\sqrt{30}/\sqrt{90} \approx 0.6$. The higher dielectric anisotropy of FNLC919 allows one to use relatively weak driving fields $(3-21)$ V/μm as compared to $(3-35)$ V/μm in the case of GPDA200 to achieve the same $\delta n_{max}$. The increase of $\delta n_{max}$ with the temperature, the inset of Fig. 3(d), demonstrates that the lower initial scalar order parameter yields better performance in terms of $\delta n_{max}$.

The observed fast (microseconds) switching times $\tau_{on}$ and $\tau_{off}$ represent the major advantage of the MEMOP effect over the Frederiks transition; as already discussed, the switch-off time of the Frederiks effect is orders of magnitude longer, $\tau_{off} \approx 25$ ms for cells comparable to the one explored in this work. The reason is that the MEMOP effect does not involve a macroscopic realignment of the optical axis [18, 35, 36], which is a slow process limited by viscosity, cell thickness, and surface anchoring factors. It is of interest to discuss the observed dependencies of $\tau_{on}$ and $\tau_{off}$ on the driving field and temperature.

The field dependences of $\tau_{on}(E)$ and $\tau_{off}(E)$ are slightly nonmonotonous, as the response times decrease with an increase of $E$ in the range $E = (3-12)$ V/μm and then somewhat increase in the range $E = (12-21)$ V/μm, Fig. 3(e). Note that a similar behavior of the response times was observed in a previous study of a multicomponent nematic mixture with a giant dielectric anisotropy $\Delta\varepsilon = 200$ [18]. The decrease of $\tau_{on}(E)$ and $\tau_{off}(E)$ is expected, as the result of a stronger dielectric torque in the "on" process and a higher field-induced orientational order in the "off" case; transient dielectric heating can also contribute to the shortening of the response times [37]. The increase of $\tau_{on}(E)$ and $\tau_{off}(E)$ at higher voltages is less clear. One possible mechanism is the dielectric memory, i.e., the dependence of the dielectric torque not only on the current value of the electric field but also on its past values [35, 38]. A strong applied field of a certain polarity, such as the one in Fig. 3 (a), creates a dipole moment density with a lifetime that is approximately equal to the Debye relaxation time, $\tau_D$. This "memorized" dipole moment density contributes to the dielectric torque acting on the nematic when the time of the voltage change is on the order of $\tau_D$. In the explored material, the Debye relaxation times increases from about 6 μs to 15 μs as temperature decreases form 80 °C to 70 °C, Fig. 1(a) and the voltage fronts in Fig. 3(a) are on the order of 0.1 μs or less. As a result, when the voltage pulse exhibits a sharp edge, it is perceived by the nematic as a high-frequency excitation, for which the dielectric anisotropy, if it is positive, diminishes, or even changes to negative. It would result in a longer $\tau_{on}(E)$. In the off-process, the decreasing but still substantial electric field can couple to the memorized dipole moment density to slow down the relaxation.

As a function of increasing temperature, $\tau_{on}(T)$ decreases, which is expected since viscosity decreases, Fig. 3(f). In contrast, $\tau_{off}(T)$ is nonmonotonous with an increase at higher temperatures. A similar nonmonotonous behavior of $\tau_{off}(T)$ has been observed for nematic mixtures with a large positive $\Delta\varepsilon$ [18]. As already suggested [18], a plausible explanation might be that in the multicomponent mixture, such as the one studied by us, nanosegregation of the components can produce different spectra of the fluctuation modes at different temperatures. Less likely is an explanation that the ratio of the relevant viscosity to the elastic constant is also increasing with the temperature since usually, the viscosity decreases faster with the temperature than the elastic moduli. Therefore, further studies are needed to better understand the nonmonotonous behavior of $\tau_{off}(T)$.

## 4. Conclusion

We demonstrate a microsecond electrically-modified order parameter (MEMOP) effect in the nematic phase of the ferroelectric nematic material FNLC919, in which the electric field on the order of $10^7$ V/m enhances the birefringence by $\delta n_{max} \approx (0.01 - 0.04)$. We stress that the effect is explored in the N phase, which allows one to construct the homeotropic cells and apply the electric field along the director by using transparent electrodes on the bounding plates. This homeotropic geometry is hard, if at all possible, to reproduce in the $N_F$ phase since the high electric polarization deposits a strong charge on the bounding plates; so far, no homeotropic alignment of the $N_F$ has been reported. Linear dependence of $\delta n_{max}$ on the field $E$ suggests that the main mechanism is the quenching of the director fluctuations. Compared to the NEMOP effect in conventional paraelectric NLCs [13, 16], the operating fields for FNLC919 are about one order of magnitude lower, and the field-induced birefringence changes are a few times stronger. The electro-optic response of FNLC919 is very fast, around 1 μs for both field-on and field-off switching.

    Overall, the FoM of the explored N material is two to three orders of magnitude higher than that of the conventional Frederiks effect [3, 32] and one order of magnitude higher than the previously reported MEMOP effect in the N material GPDA200 that does not form a ferroelectric $N_F$ phase [18]. The FoMs of the NEMOP effects in Table 1 are higher than the FoM of FNLC919, mainly because the NEMOP switching times are shorter than 0.1 μs. The 1 μs MEMOP performance of FNLC919 fills the gap between the previously reported NEMOP and 10 μs MEMOP effects. Therefore, the electro-optics of the N phase based on ferroelectric nematic materials might find applications where the response time is not required to be shorter than 1 μs. Among the electro-optical devices that could benefit from such a performance are phase modulators, optical shutters, and beam steerers, in which the propagating light beam can follow an oblique pathway as in Fig. 2(b) or even in displays if the cell is planar and addressed by an in-plane electric field.

    From the fundamental point of view, it would be of interest to explore whether the observed performance is intrinsically related to the presence of the ferroelectric nematic phase or if it can be replicated in a conventional nematic material formed by molecules with strong permanent dipole moments that do not form a ferroelectric nematic. To answer this question, one would need to prepare multiple mixtures of materials with known dipole moments and explore the phase diagram in which the ferroelectric phase disappears. A related issue is whether the electrically modified order parameter (EMOP) effects could produce enhanced performance in the ferroelectric nematic phase itself. These studies are planned for the future.

**Funding.** National Science Foundation (ECCS-2122399 and DMR-2341830).

**Acknowledgments.** We thank Sergij V. Shiyanovskii for the useful discussions. We thank Merck KGaA, Darmstadt, Germany for providing the material FNLC919.

**Disclosures.** The authors declare no conflicts of interest.

**Data availability.** Data underlying the results presented in this paper are not publicly available at this time but may be obtained from the authors upon reasonable request.